# Assessing the Impact of COVID-19 on the Objective and Analysis of Oncology Clinical Trials – Application of the Estimand Framework

Evgeny Degtyarev[a][*][#] and Kaspar Rufibach[b][#] and Yue Shentu[c][#] and Godwin Yung[d][#] and Michelle Casey[e] and Stefan Englert[f] and Feng Liu[g] and Yi Liu[h] and Oliver Sailer[i] and Jonathan Siegel[j] and Steven Sun[k] and Rui Tang[l] and Jiangxiu Zhou[m]

on behalf of the industry working group on estimands in oncology[z]

[a]Novartis Pharma AG, Basel, Switzerland; [b]F. Hoffmann-La Roche, Basel, Switzerland; [c]Merck & Co., Inc, Rahway, New Jersey, United States; [d]Takeda Pharmaceuticals, Cambridge, Massachusetts, United States; [e]Pfizer, Collegeville, Pennsylvania, United States; [f]AbbVie Deutschland GmbH & Co KG, Ludwigshafen, Germany; [g]AstraZeneca, Gaithersburg, Maryland, United States; [h]Nektar Therapeutics, San Francisco, California, United States; [i]Boehringer Ingelheim Pharma GmbH & Co. KG, Biberach, Germany; [j]Bayer US Inc., Whippany, New Jersey, United States; [k]Janssen R&D, Raritan, New Jersey, United States; [l]Servier Pharmaceuticals, Boston, Massachusetts, United States; [m]GlaxoSmithKline, Collegeville, Pennsylvania, United States

[z]Which is both, a European special interest group 'Estimands in oncology', sponsored by PSI and EFSPI, and a scientific working group 'Estimands in oncology' of the ASA Biopharmaceutical Section. http://www.oncoestimand.org

*corresponding author: evgeny.degtyarev@novartis.com

[#]these authors all contributed equally

# Assessing the Impact of COVID-19 on the Clinical Trial Objective and Analysis of Oncology Clinical Trials – Application of the Estimand Framework

COVID-19 outbreak has rapidly evolved into a global pandemic. The impact of COVID-19 on patient journeys in oncology represents a new risk to interpretation of trial results and its broad applicability for future clinical practice. We identify key intercurrent events that may occur due to COVID-19 in oncology clinical trials with a focus on time-to-event endpoints and discuss considerations pertaining to the other estimand attributes introduced in the ICH E9 addendum. We propose strategies to handle COVID-19 related intercurrent events, depending on their relationship with malignancy and treatment and the interpretability of data after them. We argue that the clinical trial objective from a world without COVID-19 pandemic remains valid. The estimand framework provides a common language to discuss the impact of COVID-19 in a structured and transparent manner. This demonstrates that the applicability of the framework may even go beyond what it was initially intended for.

## 1. INTRODUCTION

Since its initial outbreak in late 2019, Coronavirus Disease 2019 (COVID-19) has rapidly evolved into a devastating, global pandemic (Huang et al. 2020; Gates 2020). As of 31 May 2020, the disease has reached over 200 countries and territories, causing more than 5.9 million infections and 360,000 deaths (WHO 2020). COVID-19 is also having a detrimental impact on patients with underlying diseases (such as cancer) and ongoing clinical trials. Some of its impacts are direct, e.g. infections and deaths (Dai et al. 2020). Others are indirect but still deeply concerning, e.g. increased demands on the health service, travel restrictions and measures of social distancing, leading to clinical site closures, treatment interruptions/discontinuations and delayed/missed trial visits (Singh and Chaturvedi 2020). In March 2020, the U.S. Food and Drug Administration and the European Medicines Agency issued separate guidance on the conduct of clinical trials during COVID-19 (FDA 2020; EMA 2020a). The call to action was clear: 1) Ensure the safety of trial participants; 2) To minimize the risks to trial integrity and maintain compliance with good clinical practice, trial sponsors should document changes in trial conduct due to COVID-19, duration of those changes, and how trial conduct and results were impacted.

While subject safety is of greatest importance, this article is concerned with the risks COVID-19 poses to interpretability of trial results with regard to the clinical trial objective and measures to curb those risks. We argue that the clinical trial objective should relate to a world without ongoing COVID-19 pandemic, specifically defined through two criteria:

1) (a) Patients do not experience severe complications due to the virus (e.g. hospitalization and death), (b) transmission and spread of the virus are limited, and (c) effective therapy for the virus is available;
2) No major disruption of healthcare systems, patients have access to medications, routine standard of care, and proper disease follow-up.

Our interest in this context is motivated by two assumptions. First, that although not explicitly stated in protocols, clinical trials started before COVID-19 were designed with the intention to inform clinical practice in a world absent of the pandemic as defined above. Second, that this pandemic will eventually come to end, so that the clinical trial objective most relevant to patients in a post-pandemic world is the same as the clinical trial objective conceived prior to the pandemic.

To assess the impact of COVID-19 on ongoing clinical trials, the ICH E9 (R1) addendum on estimands and sensitivity analysis provides a helpful framework for discussion (EMA 2020b). Under this framework, clinical trials provide a precise description (a.k.a. the estimand or "target of estimation") of the treatment effect reflecting the clinical question posed by a given trial objective. Each estimand is made up of five attributes (Figure 1). Together, the attributes determine how different patient journeys in a clinical trial are accounted for, what data to collect, and what statistical method to use to estimate the treatment effect. Sensitivity analyses may be pre-specified to explore the estimator's robustness of inference to model assumptions (Degtyarev et al. 2019).

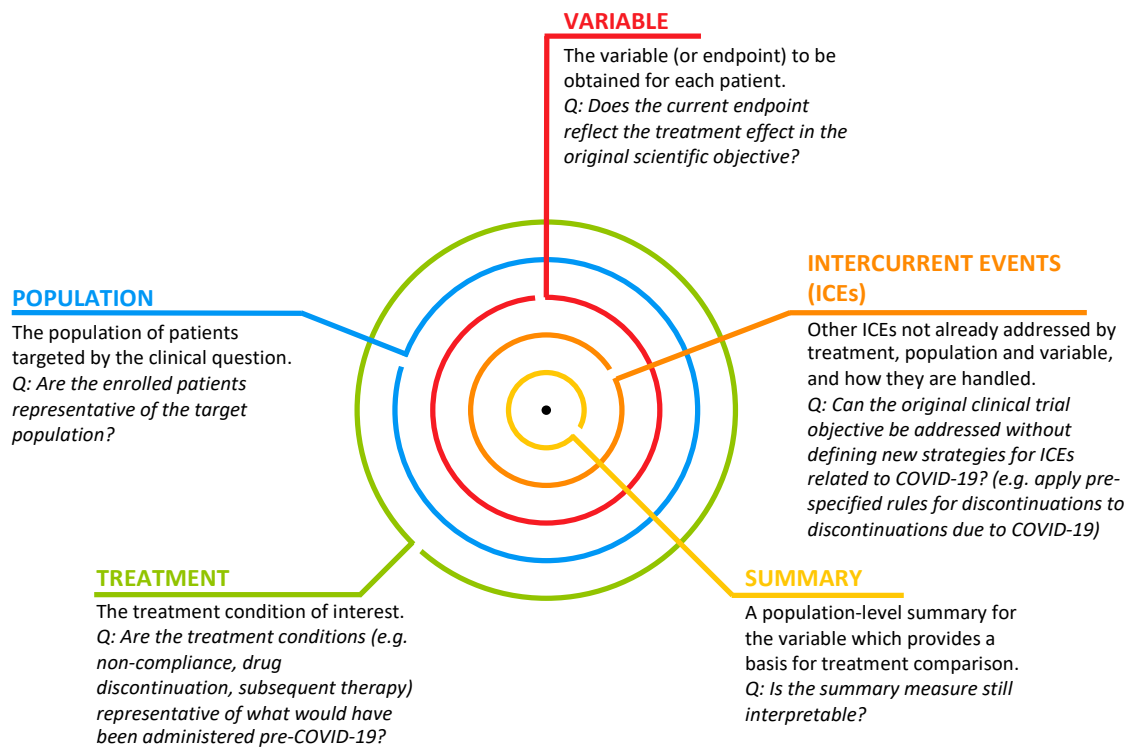

Figure 1. Covid-19 Impact Assessment. An estimand or "target of estimation" consists of five attributes: treatment, population, variable, (other) intercurrent events, and summary. We pose a key question for each attribute to facilitate COVID-19 impact assessment on whether the planned analysis of an ongoing clinical trial can still address the original clinical trial objective (in italics).

The primary intention of the ICH E9 addendum is to promote alignment between clinical trial objectives and treatment effect estimation *prior to* the start of a trial. However, it is also specific on how to handle intercurrent events (ICEs) that were unforeseen and, more generally, any changes to the estimand:

> *Addressing intercurrent events that were not foreseen at the design stage, and are identified during the conduct of the trial, should discuss not only the choices made for the analysis, but the effect on the estimand, i.e. on the description of the treatment effect that is being estimated, and the interpretation of the trial results. A change to the estimand should usually be reflected through amendment to the protocol.*

This makes the estimand framework particularly useful during the COVID-19 pandemic—an unusual, unforeseen event that has the potential to introduce ICEs and other challenges at a large scale. By having a precise definition of the initial target of estimation, sponsors of ongoing clinical trials can better identify potential sources of bias due to the pandemic. They can then answer the following questions with clarity and objectivity: "Can the estimate from my initially planned analysis still provide an answer to my clinical trial objective?" and "Will the data collected and trial results be useful for informing clinical practice in a world without COVID-19 pandemic?" If the answer to at least one of these questions is no, mitigative measures may need to be taken, e.g. by clarifying the primary estimand, modifying the estimator, adding sensitivity analyses, or introducing supplementary estimands. Ratitch et al. (2019) discuss further exploratory analyses before unblinding that may help to provide supplementary evidence for interpretation of trial results.

Recognizing the useful role that estimands may play in analysing clinical trial data affected by COVID-19, the industry working group on estimands in oncology convened in March 2020 to share experiences in dealing with emerging challenges to ongoing clinical trials (Degtyarev and Rufibach 2020). This article provides a summary of our month-long discussion and represents a consensus opinion of the working group. Given the novelty and evolving nature of the COVID-19 pandemic, we acknowledge that our opinion may need refining and enhancement in positioning over time.

Our intention in this article is to support discussions between various stakeholders (sponsors, investigators, Institutional Review Boards/Independent Ethics Committees, and health authorities) regarding the impact of COVID-19 on individual clinical trials. We assume throughout that the reader is familiar with the estimand framework as described in the ICH E9 addendum. While our focus will be on the impact of COVID-19 on randomized superiority clinical trials (RCT) in oncology, we comment on other settings and hope that this discussion will be informative for future public health crises, should there be more that cause similar disruptions to health care systems. We also refer the reader to Meyer et al. (2020) for a comprehensive treatment of other statistical and operational considerations for trials during COVID-19 in addition to their independent interpretations on estimands. While many of their advice are applicable in general, this article provides a deeper dive into oncology-specific issues from the perspective of the estimand framework.

The remainder of this article is organized as follows. In Section 2, we assess the potential impact of COVID-19 on interpretation of randomized superiority clinical trials with a focus on time-to-event endpoints. In Section 3, we discuss options for either updating the initially intended analysis or adding a supplementary estimand. Considerations for other types of endpoints, missing data, and trials with interim analyses are provided in Section 4. We conclude with a discussion in Section 5.

## 2. ASSESSING THE POTENTIAL IMPACT OF COVID-19 ON TRIALS WITH TIME-TO-EVENT ENDPOINTS USING THE ESTIMAND FRAMEWORK

### 2.1 Clinical Trial Objective

The estimand framework helps to structure the assessment of COVID-19 impact on the interpretability of trial results. We provide in Figure 1 five key questions (one for each of the five estimand attributes) that can guide trial team discussions to assess whether the estimate from the initially planned analysis can still provide an informative answer to the original clinical trial objective. If the answers to these questions indicate that interpretation of the trial may be affected by COVID-19, then depending on the anticipated magnitude of the impact, the primary estimand may need to be clarified or a supplementary estimand may need to be added. Note that changes to the primary estimand may have implications for the assumed underlying effect size, the sample size, and how missing data are handled during estimation.

Subsequently, we will describe various aspects to consider when answering the five questions in Figure 1. Prior to that, it may be useful to categorize intercurrent events as follows (EMA 2020c):

- Direct impact: caused primarily by COVID-19 infection, possibly resulting in treatment interruption or discontinuation from treatment due to infection, use of additional therapies to treat COVID-19, or death.

- Indirect impact: caused primarily by overwhelmed healthcare systems or public health measures such as regional lock-downs, possibly resulting in treatment interruption or discontinuation from treatment due to logistic reasons, patient or physician decision.

These impacts will certainly vary by regions. A non-exhaustive list of various ICEs is summarized in Table 1. In practice, distinguishing between ICEs due to direct or indirect impact of COVID-19 will require detailed data collection beyond what was planned before the pandemic. As in the ICH E9 addendum, we do not consider discontinuation from trial to be an ICE, though it may lead to missing data (for more details, see Section 4).

Table 1. Impact of COVID-19 on patient's journeys in the trial

| Type of impact | What could happen? | When more likely to happen? | Intercurrent event due to COVID-19 | Considerations for the choice of estimand strategy |
|---|---|---|---|---|
| Direct | COVID-19 infection | Risk of infection generally assumed to be equal for all enrolled patients (although it may depend on recruiting countries), however risk of severe illness as consequence of COVID-19 infection likely to be higher in patients with blood cancers[a], comorbidities or those receiving treatment associated with immunosuppression. | Death attributed to COVID-19 | Careful assessment of potential association of COVID-19 deaths and discontinuations with prognosis and trial treatment required. Treatment policy may be reasonable if association with disease progression or effect of treatment cannot be excluded. |
| | | | Treatment discontinuation due to adverse event (COVID-19 infection) | |
| | | | Use of concomitant medication to treat COVID-19 | Concomitant medications that are usually necessitated by worsening COVID-19 symptoms, which in turn may be associated with comorbidity and cancer prognosis. Furthermore, the possible impact of such concomitant medications on the disease and potential drug-drug interaction need to be considered as some anticancer therapies are currently studied as potential treatment for COVID-19 in clinical trials (U.S. National Library of Medicine 2020a,b). |

| | | | Treatment interruption due to adverse event (COVID-19 infection) | Assessment of association with prognosis and trial treatment required. Data after such interruptions may still be informative of the treatment effect without pandemic and treatment policy may be reasonable to reflect it. |
|---|---|---|---|---|
| Direct | Increased risk of immunosupression with trial treatment | Case by case assessment of benefit-risk required as many cancer treatments are immunosuppressive | Treatment discontinuation due to physician decision | Trial treatment may put a patient at higher risk of severe consequences of the infection and treatment policy may be reasonable to reflect it. |
| Indirect | Oral medication available in the target indication[b] | In trials with IV treatment requiring hospital visits | Treatment discontinuation due to patient or physician decision | Increased number of such treatment discontinuations due to desire to minimize travelling during the pandemic. Data after such discontinuations unlikely to reflect patient journeys in the post-pandemic world – hypothetical strategy could be considered. |
| Indirect | Patient can receive SoC closer to their home | In open-label trials after randomization to SoC, in particular with reimbursed SoC | Treatment discontinuation due to patient or physician decision | |
| Indirect | Patients who achieved complete or partial response on treatment not willing to travel to receive additional treatment | Trials with treatment sequences conditional on early outcomes (e.g. trials with induction-consolidation-maintenance phases[c]); possibly adjuvant trials | Treatment discontinuation due to patient or physician decision | |
| | | | Treatment interruption/delay due to patient or physician decision | While increased number of such interruptions is unlikely to happen in the post-pandemic world, data after such interruptions may still be |

|  |  |  |  | informative of the treatment effect without pandemic dependent on the length of interruption and its impact on treatment exposure/dose intensity. If interruption/delay is short, treatment policy could be considered. Hypothetical strategy may be reasonable for interruptions/delays with significant impact on patient's exposure, dose intensity or planned treatment sequence. |
|---|---|---|---|---|
| Indirect | Logistic (e.g. hospital capacity) or drug supply/manufacturing constraints | More likely for treatments requiring ICU bed availability (e.g. CAR-T), scheduled surgeries (e.g. neo-adjuvant trials) or products with complex manufacturing | Treatment interruption/delay | Increased number of such interruptions/delays only expected in the acute phase of the pandemic and unlikely to reflect the situation in the post-pandemic world, however data after such interruptions/delays may still be informative of the treatment effect without pandemic dependent on the length of interruption/delay. If interruption/delay is short, treatment policy could be considered. Hypothetical strategy may be reasonable for long interruptions/delays with |

|  |  |  |  | significant impact on intended treatment sequence. |
|---|---|---|---|---|
| Indirect | Unforeseen changes in competitive landscape (e.g. approvals of new drugs while enrolment is on hold for an extended time due to COVID-19) | In open-label trials after randomization to SoC | Treatment discontinuation due to patient or physician decision | New therapies not anticipated to be approved during the trial at trial start may impact patient's or physician's behavior and may impact trial interpretability dependent on how data after start of new therapy is used in primary analysis. |
| [a] NOTE: "Blood cancers often directly compromise the immune system, so those patients are probably most at risk, whereas cancers such as colon cancer, breast cancer, and lung cancer do not typically cause immune suppression that is not treatment-related." R. Schilsky, chief medical officer of the American Society of Clinical Oncology (ASCO) (Burki 2020) [b] NOTE: "Some patients may be able to switch chemotherapy from IV to oral therapies, which would decrease the frequency of clinic visits but would require greater vigilance by the health care team to be sure that patients are taking their medicine correctly." (ASCO 2020) [c] NOTE: "For patients in deep remission who are receiving maintenance therapy, stopping chemotherapy may be an option." (ASCO 2020) ||||||

Following the intention-to-treat principle, the treatment policy strategy is often applied in superiority RCTs to address ICEs such as treatment discontinuations, interruptions, or the use of concomitant or rescue medication. This is because these events reflect what can happen in clinical practice when the treatment is used post-approval; the underlying assumption is that these events are related to the effect of the treatment or disease or that they are likely to be observed in practice (e.g. patients forgetting to take medication). However, the widespread, systematic disruption of the global healthcare system during the pandemic is an extreme event that is hopefully temporary. Consequently, the treatment policy strategy may not match the clinical trial objective when considering discontinuations or interruptions attributed to the disruption of clinical trial operations. Additionally, patients may discontinue or pause treatment as a result of COVID-19 infection or the use of concomitant medication to treat it. Disentangling the COVID-19 effect from underlying treatment and disease may be difficult. A careful assessment is required to determine if a change in the primary estimand or additions of supplementary estimands are warranted and if so, to select the appropriate methodology. These aspects will be discussed now.

**2.2 Population**

Consistent with the original clinical trial objective, the target population should remain as originally defined by the protocol prior to COVID-19 pandemic. However, it is important to assess whether COVID-19 will impact the actual enrolled patients in the trial, and if so, whether the enrolled patients can still be considered representative of the target population. More specifically, characteristics of patients enrolled before, during and after the pandemic may be systematically different (EMA 2020c). Bias with respect to the pre-pandemic estimand may be introduced, for example, if enrolment continues only in some regions with non-severe disruption of the healthcare system. Current clinical guidance on the management of cancer patients during COVID-19 also suggest prioritization strategies based on patients' conditions, e.g. patients with compromised immune systems, elderly patients at high risk (ASCO 2020; Burki 2020; ESMO 2020). A third example is the concern that delayed diagnosis during the pandemic due to limited healthcare access may result in worse prognosis of cancer patients enrolled after the pandemic. All these complications may pose challenges to interpretation and external validity of trial results with regard to the target population.

**2.3 Treatment**

Consistent with the original clinical trial objective, the treatment conditions to be compared should remain as originally defined by the protocol prior to COVID-19 pandemic. The risk assessment of the pandemic's impact on the treatment conditions of interest will mainly focus on whether frequency and/or duration of non-compliance or drug discontinuations is above what one would have expected in pre-pandemic clinical practice. Various reasons for possible treatment delay, interruption, or discontinuation during the pandemic are summarized in Table 1 and will be discussed further below under ICEs. Furthermore, careful assessment is required in light of new anticancer therapies after discontinuation, as many oncology trials assess the treatment effect of the investigational treatment followed by any subsequent therapies (i.e. counting events observed after start of a new therapy based on the treatment policy strategy) or consider start of new therapy itself informative for the treatment effect (i.e. counting start of a new therapy as an event for the variable based on the composite variable strategy). Apart from higher number of subsequent therapies, different type of therapies compared to the pre-pandemic world may be expected during the pandemic. Therefore, the treatment sequence observed during the pandemic for some patients in the trial may not be representative of clinical practice pre- and post-pandemic.

**2.4 Variable**

Overall survival (OS) is the gold standard in oncology trials as a measure of therapeutic benefit. Other endpoints such as progression-free survival (PFS), recurrence- or event-free survival (EFS) and objective response are typically associated with tumour kinetics or disease-free status. The impact of COVID-19 related ICEs and missing data on the variable definition will be discussed in subsequent sections.

**2.5 Intercurrent Events**

Identifying ICEs and specifying strategies to handle each event are crucial components of the estimand definition in the ICH E9 addendum. Different ICEs in the same trial may require different strategies in order to address the clinical trial objective. For the remainder of this section, we will discuss in detail three potential ICEs: "death due to COVID-19", "discontinuation from treatment due to COVID-19 infection" and "discontinuation from treatment due to the pandemic but not related to COVID-19 infection". Additional considerations related to other ICEs, such as "treatment interruption", "delay of scheduled intervention" (e.g. in settings with sequences of interventions including surgery or transplant and leading to deviation from intended treatment strategy) and "use of concomitant medication to treat COVID-19", are summarized in Table 1. Considerations related to missing data (e.g. caused by withdrawal from trial) are discussed in Section 3.

   <u>Death due to COVID-19 (Direct impact).</u> In oncology trials, all-causality deaths are often counted as events for death-related endpoints, e.g. OS, PFS, and EFS. Counting a COVID-19 related fatality as an event implies that the composite strategy is used for this ICE. If we do not expect COVID-19 related deaths in a post-pandemic world, then the hypothetical strategy would be more appropriately aligned with our original clinical trial objective and would require a change in endpoint definition. Intuitively, this strategy answers the question, "What would the benefit be if patients do not die from COVID-19?" However, it may be difficult to determine whether a death is entirely attributable to COVID-19, particularly given that classifying deaths into related and not related to COVID-19 is handled heterogeneously globally (Shet et al. 2020). Even if COVID-19 is the main cause of death, a composite strategy may still be reasonable if the underlying cancer contributed to the death to some extent. Furthermore, if the number of deaths due to COVID-19 is low, then addressing these deaths using a composite strategy may approximate a hypothetical strategy with sufficient precision.

Careful assessment of the potential association of COVID-19 deaths with underlying disease prognosis and trial treatment is required before choosing the strategy and the corresponding analysis method, see Section 3.

<u>Discontinuation from treatment due to COVID-19 infection (Direct impact).</u> Treatment discontinuations due to COVID-19 medical reasons may be associated with underlying conditions or comorbidities (including cancer) or related to immunosuppressive effect of the treatment. Careful clinical assessment as well as external evidence are needed to understand whether these ICEs are associated with worse outcomes that would have been observed even had the COVID-19 pandemic not occurred. Like other discontinuations due to adverse events, it may be difficult to exclude its potential relationship to disease or effect of the treatment. Moreover, COVID-19 infection may remain a public risk for a longer time period compared to healthcare system disruption. Therefore, if treatment policy strategy is used for other adverse event related treatment discontinuations, the same strategy may be reasonable for discontinuations due to COVID-19 infection. Endpoint assessments after such discontinuations would then need to be collected for use in the analysis.

If treatment discontinuations due to COVID-19 infections are not expected in a post-pandemic world and are deemed unrelated to disease or treatment, then the hypothetical strategy may be more appropriate. However, as with deaths, the analysis corresponding to treatment policy strategy may still address the question related to hypothetical strategy with sufficient precision, if the number of such discontinuations is low.

Ideally, in the case where the number of COVID-19 infection within the clinical trial participants warrants further exploration, one could be interested to explore the difference in the effect of treatment in those infected by COVID-19 vs. those not infected (EMA 2020c). A question of potential clinical interest could be "What is the treatment effect in patients who would never experience severe complications from COVID-19 infection, regardless of what treatment they receive?" The answer to this question could provide additional insight into treatment efficacy in a post-pandemic world where effective therapy is available, in particular if the proportions of patients with such complications differ between the two arms. Such an approach defining a question of interest based on the potential of having (or in this case, not having) certain ICEs under both interventions is called *principal stratum* in the estimand framework, and in contrary to Meyer et al. (2020) we see potential usefulness of this strategy in our context. Patients in a stratum represent a subgroup of the target population which is fixed at baseline. Although in general it is not possible to identify which patients belong to each stratum, it is still possible to estimate the treatment effect of interest (more details in Section 3).

Discontinuation from treatment for reasons not related to COVID-19 infection (Indirect impact). Treatment discontinuations by patients' or physicians' decision during a pandemic may increase due to disruption of the healthcare system. Additionally, patients' behaviours may be impacted by the desire to reduce travelling in light of the risk of COVID-19 infection, resulting in discontinuation of trial treatment. Several potential scenarios are described in Table 1. These discontinuations are not expected to occur in a post-pandemic world but are likely to occur during the pandemic regardless of the disease status or treatment compared to discontinuations related due to COVID-19 infection.

Oncology trial protocols often apply treatment policy strategy for discontinuations irrespective of the underlying reason, and consider data after discontinuation as informative and relevant for the treatment effect of interest. However, an implicit assumption of the original clinical trial objective is that there is no systematic disruption of healthcare system. Hence, this approach may not accurately address the original clinical trial objective if data after discontinuation is unlikely to reflect patient's journey in a world without COVID-19 pandemic. Moreover, the interpretation is further complicated by the start of new anticancer therapy after discontinuation, since start of new anticancer therapy constitutes an event in many settings (e.g. EFS definition in hematology).

Therefore, in line with the original clinical trial objective, the hypothetical strategy appears to be more appropriate for these ICEs and would require a change in analysis. Analysis considerations for hypothetical strategy will be discussed in the Section 3. However, if the number of such discontinuations is low, then addressing the ICEs using the treatment policy strategy may provide an approximate answer to the original clinical trial objective.

## 2.6 Population-level Summary

Typical population-level summaries for time-to-event endpoints in oncology clinical trials are the hazard ratio and survival probability. The interpretability of these summary measures and assumptions of the corresponding estimation methods need to be assessed in light of the pandemic. This will be discussed further in Section 3.

## 3. ANALYSIS CONSIDERATIONS

### 3.1 Estimation Based on the Chosen Strategy for ICEs

In Section 2, we discussed use of the treatment policy, hypothetical, composite or principal stratum strategy for ICEs introduced by the pandemic. Analysis considerations for these strategies are summarized in Table 2.

Table 2. Analysis considerations for different strategies to handle ICEs

| Chosen intercurrent event strategy | Analysis considerations for time-to-event endpoints |
|---|---|
| Treatment policy | Events observed after the ICE (e.g. after discontinuation or interruption) are considered in the analysis, i.e. data collection of progression or death dates or the corresponding censoring dates required even after a patient experiences such an ICE |
| Composite strategy | The ICE (e.g. COVID-19 related death) is considered as event in the definition of time-to-event endpoint |
| Hypothetical | For hazard-based quantities, e.g. the hazard ratio, assuming absence of informative censoring the relative effect can be estimated through simple censoring at the ICE. If informative censoring cannot be excluded (e.g. patient is censored at the start of new therapy after discontinuation that could be attributed to disease), methods such as inverse probability of censoring weights (IPCW) accounting for that may be indicated (Robins and Finkelstein 2000; Lipkovich et al. 2016). These also allow to provide estimates of a hypothetical estimand for survival probabilities. |
| Principal stratum | Estimation of the treatment effect in the principal stratum such as "patients who would never experience severe impact of COVID-19 infection under either treatment" could be done within the potential outcomes framework. To estimate this effect assumptions will be necessary. A potential assumption that allows for estimation of principal stratum effects is principal ignorability (PI), an assumption similar to the ignorability assumption in propensity score analysis of observational data (Jo and Stuart 2009). PI assumes that, conditional on baseline confounders, the potential outcome (e.g. PFS or OS) for the treated (untreated) is independent of the potential outcome of the COVID-19 status for untreated (treated). Stated differently, once baseline covariates that may confound the relationship between COVID-19 status and the outcome variable are known, knowing the COVID-19 status of the treated (untreated) provides no further information on the outcome for the untreated (treated) and vice versa. Alternatively, Frangakis and Rubin (2002) use a joint model for estimation. This requires specifying two models, one for the outcome given the principal stratum and one for the principal stratum membership.<br><br>We emphasize that these assumptions are unverifiable from the collected data. Jo and Stuart (2009) and Stuart and Jo (2015) describe sensitivity analyses for principal ignorability when making the exclusion restriction assumption. As a reviewer pointed out, tipping point analyses can also be used to explore the extent to which inestimable quantities would need to vary in order to change the conclusion of the analysis. This would be an extension of, e.g., the methods proposed in Lou (2019) to superiority trials with time-to-event endpoints. |

For a hypothetical estimand asking "What is the treatment effect in a world where no patient would die due to COVID-19?", potential measures of interest include the ratio of hazards of the endpoint of interest (for a relative effect) and the cumulative incidence function in a competing risk framework (for an absolute effect). Although censoring of competing event "Death due to COVID-19" is potentially informative, the former can still be estimated by simple censoring, see e.g. the discussion in Unkel et al. (2019). For the latter, estimation should be based on Aalen-Johansen estimator. However, as Meyer et al. (2020) nicely discuss, the cumulative incidence function can still be difficult to interpret because patients are only at risk of COVID-19 death during the pandemic.

Hence, patients enrolled pre-, during-, and post-pandemic have different competing risk profiles over time from randomization. Meyer et al. (2020) conclude that this prevents interpretation of such analyses to be generalized to the population.

If one is interested in estimating the effect in patients infected by COVID-19 versus patients not infected by COVID-19, we caution against simple subsetting by this post-baseline variable as it will "break" randomization, i.e. validity of causal statements for these subgroups will be unclear. An alternative option allowing for a causal interpretation within the potential outcomes framework would be to estimate the treatment effect in the principal stratum of interest. As discussed in Section 2, one such potential stratum of interest could be "patients who would never experience severe complications of COVID-19 infection, regardless of what treatment they receive". Further considerations on the use of this methodology is discussed in Table 2.

**3.2 Additional Considerations**

As discussed in Section 2, it is important to carefully assess whether the enrolled patients remain representative of the target population. But while conceptually ideal, actually defining pre-, during, and post-pandemic periods may be challenging given that the timing of the pandemic's impact varies across regions. Potential approaches to defining pre-pandemic include date (1) of first reported case in Wuhan, (2) WHO declared COVID-19 a pandemic, (3) first reported case on country level or (4) region-specific social distancing measures were introduced. Post-pandemic could be defined as date (1) vaccination is released, (2) WHO declares COVID-19 pandemic over, (3) region-specific calls are made to end social distancing measures with no relevant rise in cases thereafter, or (4) our definition of a world without COVID-19 pandemic introduced in Section 1 applies again. Such dates would likely need to be specified on a

country- or even site-level. Furthermore, there would likely be time effects in the sense that treatment for COVID-19 infection would improve over time, and these improvements would reach different regions at different stages of the pandemic. Additional complexity may arise because periods for each patient need to be defined per endpoint. For instance, PFS may not be affected by COVID-19 for a patient if disease progression was observed pre-pandemic. However, death may happen after the start of the pandemic meaning that his OS evaluation may still be affected by COVID-19.

For tumour endpoints that rely on imaging technology, potential complications can arise from the pandemic mitigation measures such as alternative modality for imaging procedures, reduced frequency or delay of imaging. These issues should be examined as part of the overall assessment of trial integrity. They potentially induce alternative data structures that might require different statistical approaches than initially specified. An example would be delay of imaging assessments, meaning that the interval between tumour assessments could become (much) larger than specified in the protocol assessment schedule. Although in theory indicated even pre-pandemic but rarely used for such type of data, methods that appropriately deal with interval-censoring may become more appropriate (Sun 2006).

Typically, treatment effects on time-to-event endpoints are quantified using the hazard ratio. However, the data structure induced through the pandemic might imply that the proportional hazards assumption, even if it was plausible prior to the pandemic, is no longer plausible. Reasons can be that non-proportionality of hazards (NPH) is introduced due to unforeseen subgroups with similar survival in both arms (e.g. patients discontinue investigational IV treatment and switch to oral medication with similar efficacy as control arm), or competing risks with different hazard ratios (Kay 1986) (e.g. death in both arms due to COVID-19 but unrelated to treatment). As has been argued elsewhere, in the case of NPH no single summary measure can adequately capture the treatment effect entirely and alternative measures might need to be considered (Royston and Parmar 2020). Examples are difference of survival functions at a milestone or restricted-mean survival time. However, it is recognized that these alternative measures of efficacy pose their own challenges (Freidlin and Korn 2019). A comparison to the HR in superiority trials with a time-to-event end point is provided in Huang and Kuan (2018) and Yung and Liu (2019).

Whether the hypothesis test that decides a trial's success needs to be directly connected to a measure that quantifies the treatment effect remains a matter of debate (Rufibach 2019). Alternative tests that optimize power under specific scenarios for the underlying survival functions have been proposed, see e.g. Lin et al. (2020) or Royston and Parmar (2020). On the other hand, the logrank test remains valid under NPH at the cost of reduced power. As Freidlin and Korn (2019) argue, the logrank test is quite robust over a broad range of NPH scenarios, so that power loss may be compensated through a modest increase in necessary number of events (~10%) which may require increased follow-up. So, whether to change the primary hypothesis test specified pre-pandemic also requires careful consideration. Simulations might help in this assessment. During the pandemic, missing data may be induced due to travel restrictions and disruptions of the healthcare system. Specific to oncology, multiple tumour imaging scans may be missing before death/cancer progression. The importance of capturing details explaining the basis of the missing data and reporting it in the clinical trial report was highlighted in regulatory guidelines on COVID-19 impact (FDA 2020). Accurate collection of reasons for missing data, although challenging in practice, would further facilitate the exploration of the mechanism of missing data.

All the above considerations are primarily conceptual and can be done by looking at blinded data. For trials with a planned interim analysis (within a group-sequential or adaptive design), we recommend sponsors to carefully consider whether their targeted estimand and analysis strategy need to be updated. This might imply an update to the planned sample size or the targeted number of events (in the case of a time-to-event primary endpoint), e.g. because additional ICEs due to the pandemic may be reducing the targeted effect size. This also applies to priors specified in a pre-pandemic world in designs using Bayesian methods, e.g. dynamic borrowing (Viele et al. 2014).

Some of the aspects discussed above might become even more pronounced for interim analysis decisions. For example, computation of conditional power to inform a futility interim analysis depends on the observed effect size at the interim analysis together with an assumption on the effect size after the interim. For the latter, it is typically recommended to use the initially assumed effect size (Bauer and Koenig 2006). However, the initial assumption on the targeted effect size might need to be reconsidered in light of the pandemic, implying an update to the conditional power setup as well. Similarly, as discussed in Section 2, characteristics of patients enrolled before, during and after the pandemic may be systematically different, i.e. the overall trial population may be more heterogeneous compared to a world without COVID-19 pandemic. Generally speaking, for trials in which a recommendation is made at an interim analysis to the sponsor by an independent data monitoring committee (iDMC), it is paramount that the sponsor transparently informs the iDMC about the potential implications of the pandemic, changes to the estimand and analysis strategies that have been implemented. The primary purpose of an iDMC, namely to issue recommendations on safety of patients and interim analyses for trials, remains unchanged by the pandemic.

# 4. FURTHER CONSIDERATIONS FOR OTHER ENDPOINTS AND TRIAL TYPES

We discussed in previous sections the impact of COVID-19 on time-to-event endpoints. Objective response rate (ORR) is another important radiological endpoint in oncology trials. Response is usually observed at one of the first assessments in the trial, in which case ICEs due to COVID-19 or missing data at later timepoints are less likely to affect ORR evaluation. Generally, less missing data may be expected at first assessment considering its importance to determine the course of treatment and disease status for the patient. Moreover, if responses are durable, then it could still be observed at future assessments even if the first assessment is missing. More careful assessment of missing data mechanisms may be required for trials requiring confirmation of response, as it may not be captured if scans are missing before later progression. Deaths due to COVID-19 prior to an observed response would be typically considered non-responder (composite strategy as for all other early deaths) but similar considerations as described in Section 2 apply. No ideal approach for estimation of a hypothetical estimand for ORR seems to be available, but it could be done e.g. by imputing response / non-response for patients who died due to COVID-19 with a probability estimated from the proportion of responders among appropriately defined "similar patients" who had evaluable data. Alternatively, if the number of early deaths due to COVID-19 is low, an approximate estimate of a hypothetical estimand could be obtained by excluding such patients or just counting them as non-responders.

Duration of response (DOR) is more likely to be affected than ORR due to lower exposure caused by interruptions and discontinuations for various reasons (see Table 1), though responses may last far beyond treatment discontinuations with innovative therapies. Missing data may be more likely for patients in deep remission due to patient or physician decision to avoid travelling and risk of COVID-19 infection.

Careful discussion of direct and indirect impact of COVID-19 on the evaluation of DOR is needed and similar considerations for the choice of estimand strategy and the analysis as previously described could be used for this time-to-event endpoint.

In general, endpoints like ORR and DOR are often assessed in single-arm trials, for which the primary goal is not estimation of a relative treatment effect as in an RCT, but rather an absolute estimate of the quantity of interest. This estimate is then compared to historical control data that was typically collected pre-pandemic. A hypothetical strategy for intercurrent events may therefore be considered to achieve a comparison of treatment and control in a world without COVID-19 pandemic.

We agree with others who have argued that the estimand of a clinical trial does not change whether it targets superiority or non-inferiority (Akacha et al. 2017). However, while use of a treatment policy strategy in a superiority trial may reduce power (i.e. it may be conservative since efficacy is harder to establish), the same strategy used in a non-inferiority setting may inflate type I error. We recommend a careful assessment whether and how the considerations above for superiority trials apply in this setting.

Trial teams should also consider potential implications on other endpoints. Careful assessment of COVID-19 impact on safety analyses is required to ensure that trial results reflect the true safety profile of the study drug and, for example, the proportion of patients experiencing infection on trial treatment is not overestimated due to COVID-19 events. Such events might be declared as competing and estimators other than the simple incidence proportion, e.g. the Aalen-Johansen estimator, may potentially give more accurate estimates of the absolute AE risk of the actual treatment (Aalen and Johansen 1978; Stegherr et al. 2019), with the same limitation of patients potentially only being at risk during the pandemic, see Section 3.2. In particular, in the absence of randomization in single-arm trials or, if there is an overlap between known risks of the treatment and COVID-19 related AEs, the assessment of trial treatment's or COVID-19 contribution may be challenging.

## 5. DISCUSSION

In this paper, we have illustrated how the estimand framework can help to structure discussions about the impact of COVID-19 on the interpretation of results from ongoing oncology clinical trials, with a focus on time-to-event endpoints in randomized superiority trials. Other considerations that were not described in this paper can be found in a slide deck compiled by the authors at the website of the working group (Degtyarev and Rufibach 2020).

We argue that clinical trial objectives should relate to a world without COVID-19 pandemic, which implicitly assumes no major disruption of healthcare systems and absence of a highly infectious disease with severe complications and for which no effective therapy is available. Careful evaluation is required to understand whether the estimate from an initially planned analysis will accurately address this objective. If the estimate is likely to be biased in light of unforeseen COVID-19 impact, the researcher should consider clarifying the estimand, modifying the estimator, or introducing a new supplementary estimand or sensitivity analysis.

We used the estimand framework to identify several sources of potential bias. Considering various scenarios described in Section 2.2, it seems rather important to assess whether enrolled patients represent the target population. Summaries of key baseline and disease characteristics by pre-, during- and post-pandemic could be used to assess the risk of this bias, although the definition of such periods is challenging and careful interpretation of potential differences is needed, particularly in smaller trials.

We recommend that trial teams carefully assess the likelihood of treatment discontinuations and interruptions and its impact on observed treatment (or treatment sequence) in the trial. Some considerations for this assessment are provided in Section 2.5 and in Table 1. Bias may be introduced if the administered treatment regimen (including interruptions, delays, concomitant medications) in a trial is no longer representative of the intended treatment regimen. Two aspects appear to be particularly important when determining whether using treatment policy strategy to handle these ICEs is still meaningful: the potential relationship with disease progression and any effect of the treatment (e.g. immunosuppression), and interpretability of the data after the COVID-19 related ICEs, including if a patient starts new therapy or delays an intervention (e.g. surgery, transplant). For ICEs primarily caused by the disruption of healthcare systems or patients' desire to minimize travelling independently of disease or treatment, we suggest that the hypothetical strategy may be reasonable to address the clinical trial objective. Treatment policy strategy may be more appropriate for discontinuations, interruptions or delays related to COVID-19 infection. Principal stratification assessing the treatment effect in e.g. patients who would not experience severe complications of COVID-19 on either treatment arm may be considered as another alternative to provide insight on expected treatment benefit. The potential use of different strategies for treatment discontinuations further emphasizes the importance of appropriate data collection as highlighted in regulatory guidelines.

Trial-specific discussions between sponsors and regulators would be required prior to implementing clarifications to the primary estimand in a protocol amendment. In general, the impact of all these potential changes to the primary estimand may need to be explored in simulations, requiring assumptions on anticipated effects of COVID-19 on estimand attributes. In addition to clarification of the estimand, changes in the sample size and/or trial duration to allow for the observation of additional events may need to be considered. For example, longer follow-up time to observe the planned number of events may be expected in studies with censoring of new anticancer therapies due to higher number of discontinuations due to COVID-19 followed by the initiation of subsequent therapies.

In practice, dependent on the stage of the trial and the impact of COVID-19, the initially planned analysis may still provide a sufficiently precise answer. Hence, we foresee no change in primary analysis for most trials due to pragmatic considerations. Supplementary analyses accounting for ICEs due to COVID-19 in different ways or sensitivity analyses exploring different assumptions for missing data could be described in an amendment to the statistical analysis plan.

In conclusion, the impact of COVID-19 on patient journeys in oncology represents a new risk to interpretation of trial results and its usefulness for future clinical practice. Further discussions and proper data collection are needed to be able to fully assess and mitigate its impact. However, it is already fair to say that the estimand framework provides various stakeholders a common language to discuss the impact of COVID-19 in a structured and transparent manner.

# ACKNOWLEDGMENTS

We thank numerous colleagues in our companies who influenced our thinking on the presented topics. We are grateful to the editors for inviting us to contribute a paper to this special issue.